\documentclass[a4paper]{jpconf}

\usepackage{graphicx}
\usepackage{amsmath,amssymb}

\usepackage{citesort}
\usepackage{setspace}

\begin{document}

%\doublespacing
%\onehalfspacing
%\singlespacing

\title{Electron rescattering at metal nanotips induced by ultrashort laser pulses}

\author{G Wachter, C Lemell, and J Burgd\"orfer}
\address{Institute for Theoretical Physics, Vienna University of Technology, Wiedner Hauptstr. 8-10, A-1040 Vienna, Austria, EU}
\ead{georg.wachter@tuwien.ac.at}
\begin{abstract}
We theoretically investigate the interaction of moderate intensity near-infrared few cycle laser pulses with nano-scale metal tips. Local field enhancement in a nanometric region around the tip apex triggers coherent electron emission on the nanometer length and femtosecond time scale. The quantum dynamics at the surface are simulated with time-dependent density functional theory (TDDFT) and interpreted based on the simple man's model. We investigate the dependence of the emitted electron spectra on the laser wavelength.
\end{abstract}

\section{Introduction}
Since the early days of quantum physics, electron emission from solid surfaces has played a key role both as a probe for structural and dynamical information about the solid state system and to study conceptual aspects of light-matter interaction. The availability of intense few-cycle laser pulses has opened up a new regime where non-linear, or time-dependent, response properties can be probed. The interaction of metal nanotips with few-cycle laser pulses has come into focus as a unique testing ground for such strong field physics of a solid-state system. A prerequisite for the observability of strong field effects without the destruction of the solid state system is the local field enhancement near the apex of the tip. The resulting confinement of electron emission in both space to a nanometric emission area and to a femtosecond time scales gives rise to the observability of coherence and sensitivity to details of the laser pulse shape in the emitted electron spectra \cite{Schenk2010StrongField,Bormann2010TipEnhanced,Kruger2011Attosecond,Wachter2012Electron,Kruger2012Interaction}. Coherence and pulse shape sensitivity are barely observable in electron emission from solid surfaces where averaging over the phase and intensity distribution of a focal spot is inevitable \cite{Lemell2003Electron,Dombi2004Direct}. One feature associated with strong field physics has been found in the high-energy part of the electron spectra which show a plateau region followed by a cut-off \cite{Kruger2011Attosecond,Wachter2012Electron}. The cut-off stems from a process known as ``rescattering'' in strong field physics for atoms: First, the electron tunnels out near the maximum of the laser field. Subsequently, it is driven back towards the parent core as the laser field changes sign, where it can elastically scatter in backwards direction from the core reaching a maximum energy of $10 U_\mathrm{p}$ with the ponderomotive potential $U_\mathrm{p} = F_0^2 / 4\omega^2$ at laser field $F_0$ and angular frequency $\omega$. This $10 U_\mathrm{p}$ cut-off has been observed experimentally in electron spectra from nano-tips \cite{Kruger2011Attosecond} and its origin has been traced unambiguously to the rescattering process \cite{Wachter2012Electron,Kruger2012Interaction}. 

Another unique feature of electron emission from nano-tips is the inhomogeneity of the enhanced near field on the length scale of the tip radius changing the classically permitted trajectories including the ones responsible for rescattering. This has been the subject of recent studies \cite{Herink2012Fielddriven,Park2012Strong}. The inhomogeneity becomes observable when the electron quiver radius $\alpha_q = F_\mathrm{eff}/\omega^2$ with $F_\mathrm{eff}$ the local field at the surface becomes comparable to the tip radius, requiring either sharp tips or long wavelengths and high intensities. In the present study we restrict ourselves to the parameter regime where the inhomogeneity of the field does not yet play an important role. The ionization and rescattering mechanism is thus not affected by the complication of a space-dependent electric field.

The plan for this paper is as follows. First, we present a typical result for the near field enhancement at a nanotip. We then focus on the quantum surface dynamics driven by the enhanced near-field. We perform simulations based on time-dependent density functional theory and interpret the simulations with a version of the simple man's model adapted for nanostructures based on Gaussian wavepackets. Finally, we study the wavelength dependence of the electron emisson and rescattering process from visible to near-infrared wavelengths. Atomic units are used unless stated otherwise.

\section{Local field enhancement} \label{sec:fieldenh}
At the heart of the observability of strong field physics at low nominal laser intensities as well as the coherence of the electron spectra lies the local field enhancement near the tip apex. In the following we present typical results for the near field enhancement and field distribution around the tip. 

\begin{figure}[h!] 
\centerline{\includegraphics[width=0.5\textwidth]{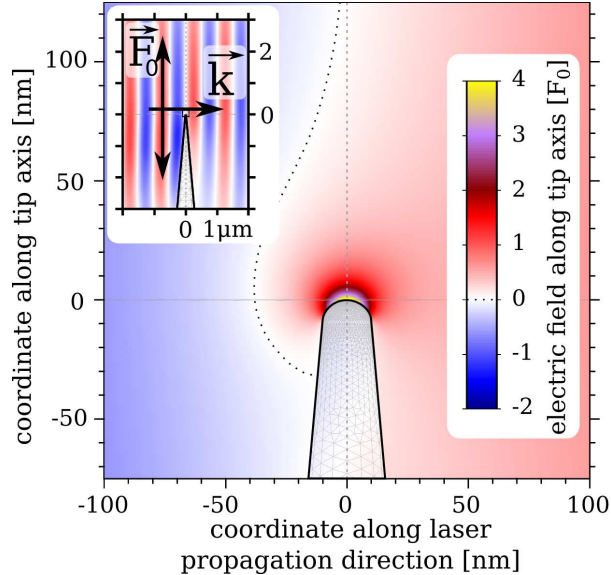}}
\caption{\label{fig:fieldenh}
Field distribution around a tungsten nanotip of 10 nm radius calculated with the boundary element method ({\sc scuff-em} package \cite{Reid2009Efficient}). The field is enhanced and phase shifted compared to the incident field. The field enhancement at the tip surface is 5.1 and the phase shift is 70 degrees.  The inset displays a zoomed-out view on the micrometer scale. The shadowing effect of the tip shank can be observed. 
% field enhancement 1 AA in front of tip apex = 1.778 + 4.771 i --> abs% = 5.09, arg%=atan2(y,x)=69.6 deg -- see gwachter@vsc:~/scuff-em/tip_W_nearfield_frq_NIR/focus/nicepic
}
\end{figure}

The Maxwell equations are solved for a laser of wavelength 1 $\mu$m linearly polarized along the tip axis with a Gaussian beam waist ($1/e$ amplitude) of 4.5 $\mu$m with the boundary element method employing the public domain package {\sc scuff-em}\cite{Reid2009Efficient}. The tungsten tip is modeled as the concatenation of a truncated sphere of radius 10 nm (the apex),  a truncated cone of half opening angle 5 degrees and length 10 $\mu$m, and another truncated sphere to provide a smooth base, with a dielectric function of $\varepsilon = -2.6 + 20.9 i$ \cite{Palik1991Handbook}. The tip surface is discretized by $\sim$ 6000 triangular panels, where the apex region is sampled to sub-nm resolution ensuring numerical convergence down to atomic resolution. Fig.~\ref{fig:fieldenh} displays a snapshot of the electric field component along the tip axis. The electric field is enhanced and phase shifted with respect to the incoming field (enhancement factor $\sim$ 5.1 and phase shift $\sim 70$ degrees), and the near field decays on the length scale of the tip radius.

\section{Electron emission from the tip apex} \label{sec:elem}
In this section, we briefly review our model \cite{Wachter2012Electron,Kruger2012Interaction,Wachter2012ElectronSPIG} of the quantum surface dynamics and electron photoemission by the near field where the locally enhanced field at the tip apex calculated in the preceding section enters as input.

We simulate electron emission with time-dependent density functional theory (TDDFT) \cite{Liebsch1997Electronic,Burke1998Density,Maitra2002Ten,Lemell2003Electron}).  Since the nano-tip apex (size a few nm) is much larger than the electron length scale (Fermi wavelength $\sim$ 0.1 nm), the symmetry in the surface plane is approximately conserved and we restrict ourselves to a one-dimensional description with the surface normal as reaction coordinate. The time-dependent electron density is expanded into Kohn-Sham orbitals $\psi_k(z,t)$
\begin{equation}
n(z,t)=\sum_{k=1}^{n_{\mathrm{occ}}} c_k |\psi_k(z,t)|^2 \label{eq2} \quad ,
\end{equation}
with the sum extending over the occupied orbitals up to the Fermi energy ($n_{\mathrm{occ}} \sim 50$ for a slab of 150 a.u.~width). The coefficients $c_k$ reflect the density of states projected on the reaction coordinate \cite{Eguiluz1984Static}. In line with earlier work, we model a tungsten tip within the jellium approximation with parameters Wigner-Seitz radius $r_{\mathrm{s}}=2.334$ a.u.~corresponding to the s-electron density of tungsten,  a Fermi energy of 9.2 eV, and a work function of $W=$ 6.2 eV. The time evolution of the electron density under the influence of an external potential $V_{\mathrm{ext}}(z,t)$ is given by the time-dependent Kohn-Sham equations
\begin{equation} \label{eq:tdks}
i \partial_t\psi_k(z,t) = \left\{-\frac{1}{2} \partial^2_z + V[n(z,t)] + V_{\mathrm{ext}}(z,t)\right\}\psi_k(z,t)\, ,
\end{equation}
which are coupled by the density-dependent potential $V[n(z,t)]$ containing the electrostatic and exchange-correlation potentials. For the ground state potential we employ a smooth jellium effective potential with the asymptotically correct $1/4z$ image potential tail to emulate the correct shape of the tunneling barrier \cite{Burgdorfer1994Atomic}.  While in the jellium model the ionic cores are smeared out to a homogenous background potential, we re-introduce the core potential of the first layer to account for rescattering employing a soft-core screened Coulomb (Yukawa) potential with the screening lenght given by the Thomas-Fermi screening length ($\sim 1$ a.u.). The simulated electron spectra are largely independent of the employed core potential for the first-layer as long as the (over-barrier) reflection coefficient is similar, as has been numerically demonstrated in \cite{Wachter2012ElectronSPIG}. The electrostatic potential $V_{es}(z,t)$ is given by the one-dimensional Poisson equation $ \frac{ d^2 }{dz^2 } V_{es} (z,t) = - 4 \pi n(z,t)$. We employ the adiabatic local density approximation in the Perdew-Zunger parametrization for the exchange-correlation potential \cite{Perdew1981Selfinteraction}. 

The external potential is given in dipole approximation as $V_{\mathrm{ext}}(z,t) = z F_{\mathrm{eff}}(t) + z F_{\mathrm{dc}}$ where the effective time-dependent laser field $F_{\mathrm{eff}}(t)$ contains the local field enhancement, and a small extraction field $F_{\mathrm{dc}}=0.2$ GV/m is adiabatically switched on before the laser pulse starts. 

We integrate the Kohn-Sham equations (\ref{eq:tdks}) by the Crank-Nicholson method in a box of size $\sim$ 2000 a.u.~(slab size $\sim 140$ a.u.) containing 13000 points and absorbing potentials to take care of unphysical reflections. The energy spectra of emitted electrons are calculated from the temporal Fourier transform of the wave functions at a large distance (1500 a.u.) from the surface \cite{Pohl2000Towards,DeGiovannini2012TextitAb,Dinh2013Calculation} so that the laser pulse is over before the electrons pass the detection point. The spectra are broadened by 0.5 eV corresponding to a typical experimental spectrometer resolution where the energy zero corresponds to the vacuum level in the absence of a static field. 

\begin{figure}[h!] 
\centerline{\includegraphics[width=1.0\textwidth]{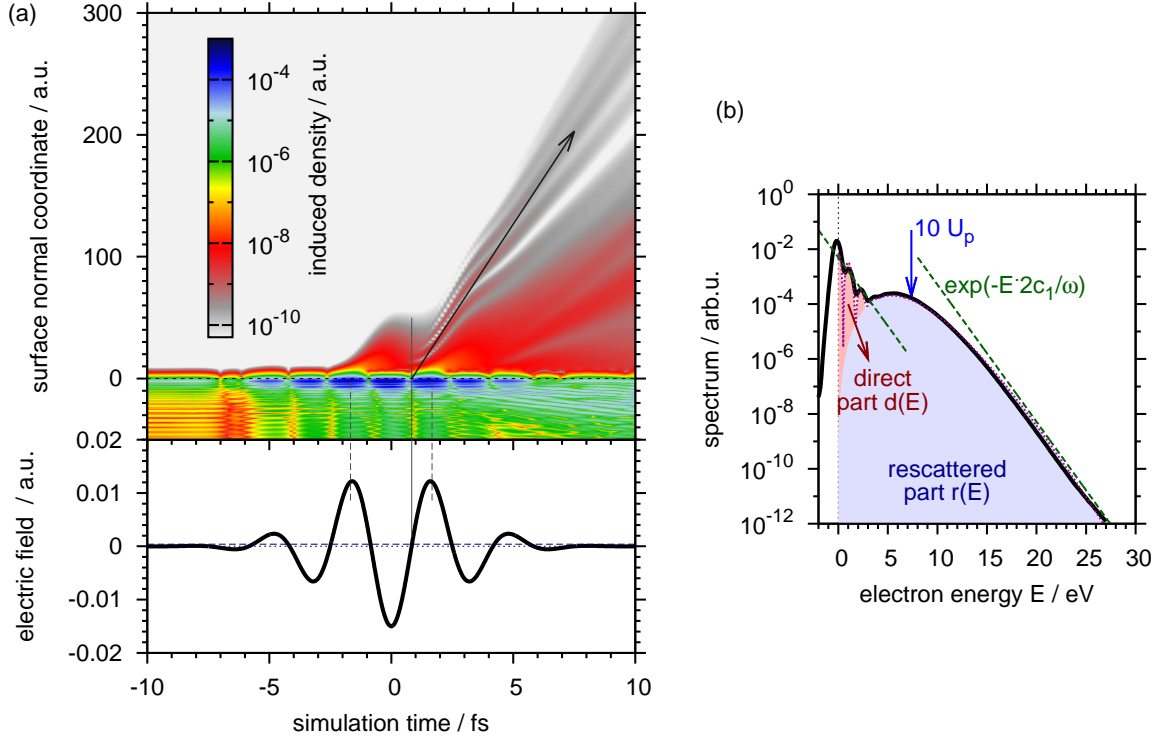}}
\caption{\label{fig:tddft}
%/home/wachter/Desktop/CN_tddft/tddft_icpeacproc/results/test_0003_F0_0.015_tau_06.00_wscan_500_1000nm/F0_0.015_Fdc_-0.20_w_0.045563_tau_06.00_phi_1.00_delay_1000.
% bounding box: opt. 55 70 410 298
Simulation of electron emission with time-dependent density functional theory (TDDFT) for an enhanced laser field strength of $F_\mathrm{eff} = 0.015$ a.u.$= 7.7$ GV/m (intensity 7.9$\times 10^{12}$ W/cm$^2$) at a central wavelength of 1 $\mu$m (photon energy $\omega=$1.23 eV) and duration 6 fs (FWHM field), a static extraction field of 0.2 GV/m, and a $-\cos$-like carrier-envelope phase. (a) Upper panel: induced density $|n(z,t) - n(z,-\infty)|$ on a logarithmic scale. The gray arrow starts at the zero of the electric field on the surface $z=0$ and its slope corresponds to the maximal classical velocity, $dz/dt = \sqrt{ 2 \times 10 U_\mathrm{p} }$. Lower panel: time-dependent electric field (solid line) and static electric field (dashed). (b) Electron spectra extracted from the simulation (black line). The cut-off of the spectra approximately coincides with $10 U_\mathrm{p}$ (blue arrow). The slope of the direct and, asymptotically, rescattered parts is approximately given by $2 c_1 / \omega = 0.92$ /eV (parallel dashed lines). The spectrum is well described by a superposition of two direct (red shaded area) and one rescattered wave packet (blue shaded area) within the simple man's model for nanostructures (eq.~\ref{eq:smmn}) \cite{Kruger2012Interaction}.
}
\end{figure}

A typical simulation (fig.~\ref{fig:tddft}, left panel) displays the time-dependent induced density response to a laser pulse with local intensity 7.9 $\times 10^{12}$ W/cm$^2$ (the region $z\le 0$ is inside the tip). Taking into account the field enhancement factor of 5.1 calculated in the preceding section, this local intensity at the tip surface can be reached with a nominal laser intensity of only $3 \times 10^{11}$ W/cm$^2$. The laser field induces a time-dependent polarization charge layer near the surface region (blue contours in fig.\ref{fig:tddft}) as electrons are moved to screen the external field. The fraction of moved electrons is small compared to the valence electron density of $n_0 = 1.88 \times 10^{-2}$ a.u., and the polarization charge layer decays within a few atomic units into the bulk, showing some Friedel oscillations with half of the Fermi wavelength ($\sim 3.8$ a.u.). Around the maxima of the electric field (dashed lines), charge is emitted towards the detector ($z>0$). The emitted charge density is, in turn, orders of magnitude lower than the polarization charge density  near the surface. As the laser field changes its sign, some of the emitted electrons are driven back towards the surface, where they can rescatter from the first atomic layer. According to the simple man's model \cite{Corkum1993Plasma}, the classical trajectories attaining eventually the maximum final energy of $\sim 10 U_\mathrm{p} = (F_\mathrm{eff}/2 \omega)^2$ start slightly after the field maximum and recollide with the surface near the zero crossing of the electric field (solid vertical line in fig.\ref{fig:tddft}(a)). The black arrow starting at the zero crossing of the electric field and with slope (velocity) corresponding to the maximum attainable energy of $10 U_\mathrm{p}$ describes the fastest outgoing electrons very well, so that we identify this portion of the time-dependent electron density with the rescattered electrons. Notably, some of the outgoing electrons are even faster than classically allowed, reflecting the quantum nature of the electron wave packet. The wave packet spreads longitudinally as it approaches the ``detector''. The second maximum of the electric field (dashed line near 2 fs) again leads to the emission of charge near the field maximum, but this time no rescattering can take place as the force towards the surface is not strong enough to drive the electrons back. The interferences visible after the second laser field maximum give rise to the multi-photon peaks observed in the direct part of the spectrum. 

The corresponding electron spectrum is shown in the right panel of fig.~\ref{fig:tddft} (solid black line). The basic features of the spectrum can be explained with the simple man's model \cite{Lewenstein1994Theory,Lewenstein1995Rings} and its recently proposed extension to nanostructures \cite{Kruger2012Interaction} in terms of Gaussian wave packets which represent an ensemble of classical trajectories contributing to the spectra. The electron momentum spectrum $P(p)$ can be parametrized as the sum of a direct part $d(p)$ and a rescattered part $r(p)$ (dashed line in fig.\ref{fig:tddft}(b))\cite{Kruger2012Interaction},
\begin{equation}  \label{eq:smmn}
P(p) = d(p) + r(p)  \quad . 
\end{equation}
The direct part (red shaded area in fig.\ref{fig:tddft}(b)) is, in turn, a superposition of two wave packets emitted by tunneling ionization around the two subsequent field maxima delayed by the laser period $T=2 \pi / \omega$ with respect to each other. Their coherent superposition leads to interferences in the energy domain (``multi-photon peaks'')
\begin{equation} 
d(p) \propto \exp(-p^2 / \sigma_p^2) \left\{ 1 + \cos\left( [p^2 /2 + U_\mathrm{p} + W] T + \phi \right) \right\} \quad, 
\end{equation}
where the phase $\phi$ is associated with the deviation of the pulse shape from a strictly sinusoidal pulse. The momentum width $\sigma_p$ is given by the longitudinal tunneling momentum distribution \cite{Popov1999Energy}, $\sigma_p^2 = \omega / 2 c_1$, $c_1(\gamma) = \mathrm{ArcSinh}(\gamma) - \gamma (1 + \gamma^2)^{-1/2}$ in terms of the Keldysh parameter $\gamma = \sqrt{2 W} \omega / F_\mathrm{tun}$ with the field strength at tunneling time $F_\mathrm{tun} = 0.83 F_\mathrm{eff}$. This distribution reduces to the ADK longitudinal momentum distribution in the deep tunneling regime $\gamma \ll 1$. As interferences in energy domain are a direct consequence of the repetition of the tunneling process in time, they are absent from the high-energy part of the spectrum which is generated by a single rescattered wave packet (black arrow in fig.~\ref{fig:tddft}(a)). The resulting dependence of the visibility of the interference fringes on details of the pulse shape (carrier-envelope phase) has been experimentally observed \cite{Kruger2011Attosecond}. The spectrum of the rescattered wave packet is given by (blue shaded area in fig.\ref{fig:tddft}(b))
\begin{equation} 
r(p) \propto \exp(- [p - p_r]^2 / \sigma^{2}_p)  \quad . 
\end{equation}
The rescattered wave packet is displaced by the gained momentum $p_r$. The order of magnitude of $p_r$ is given by the maximal achievable classical momentum $p_\mathrm{max}$ where $p_\mathrm{max}^2 / 2 = 10 \, U_\mathrm{p}$ for a harmonic laser field (blue arrow in right panel). Compared to a harmonic laser field, the maximal energy is slightly reduced by about 7 \% due to the effects of the static field and the short pulse envelope. We find best agreement to the TDDFT spectrum for an even smaller value of $p_r = 0.82 \times p_\mathrm{max}$ (maximum of the blue shaded area in fig.~\ref{fig:tddft}) which we attribute to the interplay of tunneling probability and energy gain for the ensemble of trajectories contributing to the plateau. Overall, the main features of the TDDFT spectrum are well described by the simple man's model for nanostructures. 

% While the maximum final energy is reached for trajectories starting around $0.05 T$ after the field maximum, the tunneling probability at this time is strongly decreased compared to the field maximum so that the maximum of the plateau part of the spectrum constitutes of trajectories starting somewhat earlier which, consequently, attain a smaller final energy. [sentence too complicated]

\section{Dependence on laser wavelength} \label{sec:wavelength1}
We explore now the influence of the laser wavelength on the simulated electron spectra. We vary the laser wavelength by a factor of 4 while keeping the carrier-envelope phase, field envelope, and pulse duration fixed. We note that the field enhancement and near-field phase shift as determined by the Maxwell equations are also wavelength dependent, which we, for simplicity, neglect in the following.

\begin{figure}[h!] 
\centerline{\includegraphics[width=1.00\textwidth]{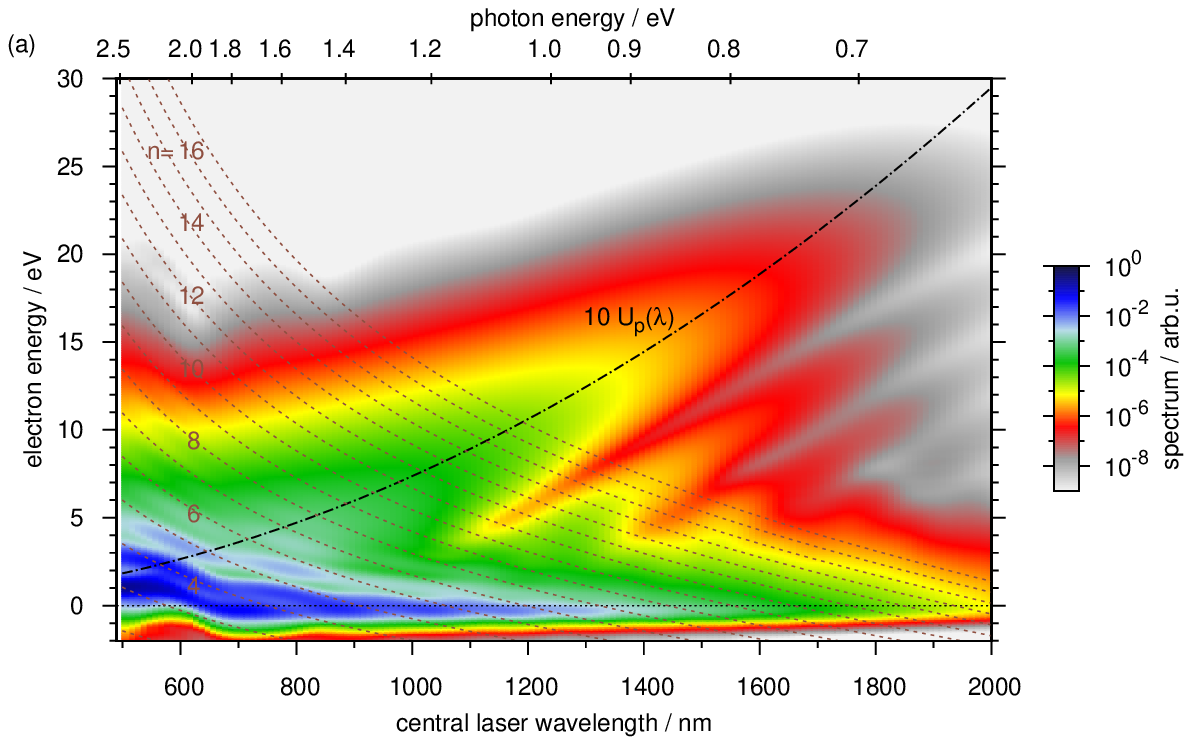}}
\centerline{\includegraphics[width=1.00\textwidth]{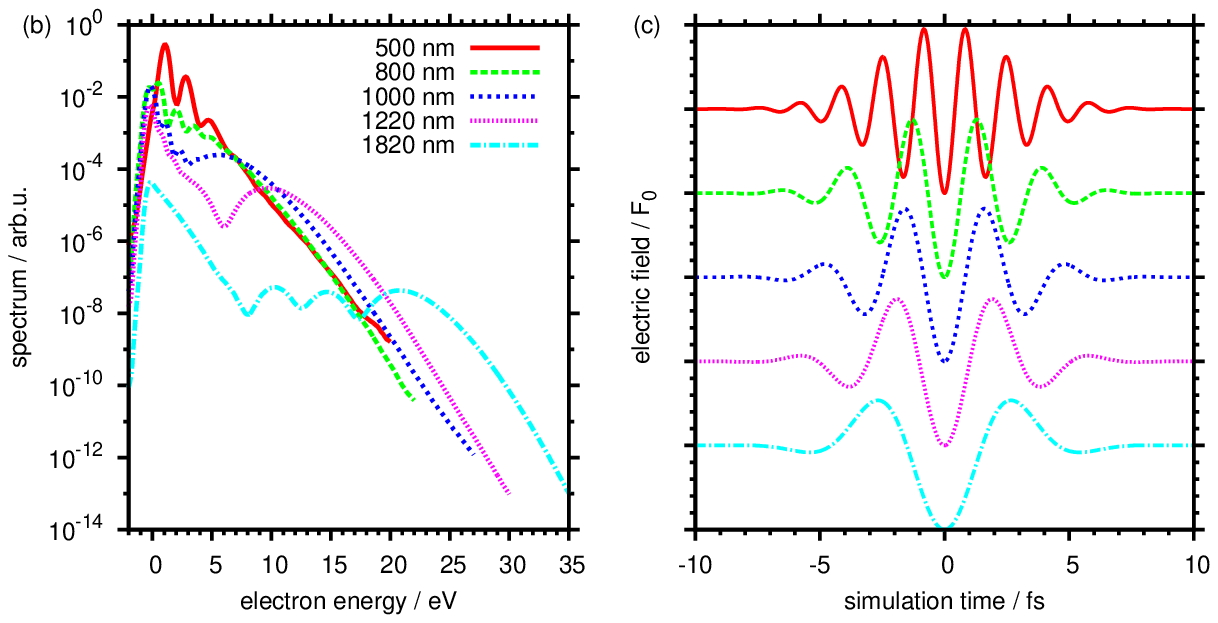}}
\caption{\label{fig:ldep}
(a) Wavelength dependence of electron emission (TDDFT simulation for the laser pulse shapes shown in (c)). Dash-dotted line marks the classical cut-off for rescattering, $10 U_\mathrm{p}(\lambda)$. Dotted lines give estimates for the position of multi-photon peaks, $E_n = n \omega - U_\mathrm{p} - W$ where $n$ is shown in the insets. (b) Line-out of (a) at various wave lengths.
}
\end{figure}

Fig.~\ref{fig:ldep} shows electron spectra simulated with TDDFT for varying laser wavelength from $\lambda = 500$ nm to $\lambda=2000$ nm with all other laser pulse parameters kept unchanged. Within this wavelenght range, the Keldysh parameter $\gamma(\lambda)\propto 1/\lambda$ varies between 4 and 1 so that we expect a transition from the multi-photon to the non-adiabatic tunneling regime with increasing wavelength. On the other hand, the classical cut-off energy for rescattering trajectories $10 U_\mathrm{p} \propto \lambda^2$ varies between 2 and 30 eV in the same wavelength range. Indeed the results from fig.~\ref{fig:ldep}(a) indicate a transition from the multiphoton regime with peaks approximately given by $E_n = n \omega - U_\mathrm{p} - W$ (dashed lines) to spectra dominated by the characteristic rescattering plateau with a cut-off energy of $10 U_\mathrm{p}$ (dash-dotted line). The intensity is reduced by several orders of magnitude from $\lambda = 500$ nm to 2000 nm as evidenced by the line-out plot in fig.~\ref{fig:ldep}(b). This decrease is partially due to lower field strength at the time of emission as the pulse length becomes comparable to the laser period (see pulse shapes given in fig.~\ref{fig:ldep}(c)). However, the increase in signal between 800 nm and 500 nm is attributed to more efficient multi-photon ionization as fewer photons are necessary to overcome the excitation gap ($W = 6.2$ eV) to the continuum. The height of the plateau region relative to the direct part strongly decreases with increasing wavelength by about 2 orders of magnitude as $\lambda$ increases from 1000 nm to 2000 nm. A minor part of this decrease is due to the decreasing reflection coefficient of our one-dimensional first layer potential by about a factor of 4 between 2 and 10 eV scattering energy $\sim 3 U_\mathrm{p}$ reached by the fastest electrons at the time of the rescattering. Partial-wave calculations performed for a three-dimensional atomic tungsten potential \cite{Salvat2005ELSEPADirac} give a somewhat smaller decrease of about a factor two in the same energy range. We speculate that the major contribution to the decrease in plateau intensity reflects the different tunneling probability near the field maximum (direct part) compared to the tunneling probability at the birth time of the rescattered wave packet (about 0.05 $T$ after the field maximum). This frequency dependence of the instantaneous tunneling rate is predicted by non-adiabatic tunneling theory \cite{Yudin2001Nonadiabatic} but has so far not been experimentally verified. Another prominent feature in our calculations is the build-up of additional large-scale structures in the spectra. We observe pronounced modulations (intensity variation by a factor $\sim 10$) spaced by several eV exceeding the photon energy. Most likely, these structures can be associated with the interference of short and long trajectories within one laser cycle (``intracycle'' interferences) similar to the structures found in ATI spectra for gas targets \cite{Paulus2002Identification,Arbo2010Intracycle,Arbo2010Diffraction,Xie2012Attosecond}. 

\section{Conclusions}
We have presented an overview of our recent progress in the study of electron emission from nano-scale metal tips by near-infrared few cycle laser pulses. The local field enhancement near the tip apex leads to strong field physics at low nominal laser intensity. Our microscopic description of the surface dynamics is based on time-dependent density functional theory (TDDFT) and can be interpreted in terms of a semi-classical wavepacket model that, despite its simplicity, retains the essential physics. We have studied the wavelength dependence of electron emission with TDDFT in terms of the transition from the multiphoton regime to the tunneling regime where the rescattering plateau dominates the spectrum. 

\section*{Acknowledgments}
This work was supported by the Austrian Science Foundation FWF under Proj.\ Nos.\ SFB-041 ViCoM, SFB-049 Next Lite, and P21141-N16. G.W.\ thanks the International Max Planck Research School of Advanced Photon Science for financial support. Calculations were performed using the Vienna Scientific Cluster (VSC). %The authors would like to thank M.\ Schenk, M.\ Kr\"uger and P.\ Hommelhoff for discussions.

%%%%%%%%%%%%%%%%%%%%%%%%%%%%%%%%%%%%%%%%%%%
\section*{References}

\bibliography{georgwachter_nourl}

\end{document}